\documentclass[12pt]{article}
\usepackage[cp1251]{inputenc}
\usepackage{epsfig}
\usepackage{graphicx}

\begin{document}

\noindent{\Large\bf Search for Keplerian periods in light variations of T Tauri stars and Herbig Ae stars}
\\
\\
\noindent{S. A. Artemenko, K. N. Grankin, P. P. Petrov}
\\
\smallskip
\noindent{\small\it{Crimean Astrophysical Observatory, Ukraine}} \\
\\
Long-term, uniform series of UBVR observations of T Tauri and Herbig Ae stars obtained over 20 yr 
at the Maidanak Observatory as part of the ROTOR program are analyzed. 
We find a linear relationship between the characteristic variability time scale and the bolometric 
luminosity of the star+disk system: the higher the luminosity, the slower the brightness variations. 
This dependence is valid over a wide range of masses and luminosities, from T Tauri stars to Herbig Ae stars. 
On average, the variability time scale is one-quarter the Keplerian period at the dust-sublimation radius, 
which is known from interferometric observations. Some T Tauri stars have periods from 25 to 120 days, 
which are preserved over several observing seasons. These periods correspond to Keplerian orbits with 
semi-major axes from 0.14 to 0.52 AU. The results obtained provide indirect evidence for the existence 
of protoplanets in the gas-dust disks of stars in early stages of their evolution toward the main sequence. 
\\
\\
{\bf 1. Introduction}
\\
\\
The young pre-main sequence objects are known as irregular variables. 
By their physical parameters they fall into two large groups: the T Tauri stars 
with masses $\leq$2.5 M$_\odot$,
and the Herbig Ae-Be stars with masses 1.5--15 M$_\odot$. These groups differ not only in 
mass and luminosity, but also in internal structure of stars: unlike Herbig Ae-Be 
stars, T Tauri stars have outer convective envelopes. Detailed description of the observed 
characteristics of T Tauri stars and Herbig Ae-Be stars can be found in review papers 
[1-3]. The young stars possess circumstellar gaseous dusty dusks, where
the process of planetary system formation is presumably going on. In recent years, the
protostellar disks have been investigated intensively by means of optical interferometry
[4]. In this paper we consider photometric variability of the Herbig Ae stars (HAeS)
and the classical T Tauri stars (cTTS) with accretion disks.

The two main sources of the irregular light variability of young stars are 1) extinction 
and scattering of light in dust clouds of circumstellar disk, and 2) accretion processes:
infall of matter on stellar surface is accompanied by energy liberation, which changes
the apparent stellar brightness. The accretion processes are more evident in cTTS of 
late spectral types, where the accretion luminosity is comparable to stellar luminosity
and sometimes even exceeds it. On the other hand, the circumstellar extinction is more 
noticeable in HAeS with more massive dusty circumstellar disks. Many HAeS shows irregular
deep minima of brightness, caused by obscuration of star by dust clouds. When a star is 
sufficiently screened out by dust, contribution of scattered light increases which results
in rise of linear polarisation and sometimes in bluer colours (UX Orionis phenomenon, [5]). 
The photospheric spectrum of such star remain unchanged. Some of cTTS of spectral types 
earlier than K2 also show the effect of UX Ori. The role of different mechanisms of
variability of cTTS was discussed in the recent paper by Grankin et al.[6]. One more
reason of photometric variability is the magnetic activity, which reveal itself,
specifically, in appearance of cool spots on stellar surface. This effect is present only
in stars with convective envelopes. Due to long life time of the cool spots, they can be
discovered from steady rotational modulation of brightness with periods of several days [7].

A protoplanetary disk is not uniform. The presence of graviting bodies in the disk result 
in regular perturbations of density, which can be detected from periodical brightness 
variations, provided a favorable orientation of the disk to the line of sight. The density 
waves can appear not only on the disk plane, but also in the disk wind [8], which increases
the probability of their detection. Herbig Ae stars seem to be the most promising objects 
to look for such periodicities. Light curves of the stars were analysed in details in [9-15],
but no stable periods were detected. At longer time intervals of months and years, wave-like
variations of the maximum level of brightness were found. In some stars a cyclic pattern of 
these variations was revealed (e.g., in SV Cep [12]). Such photometric behavior indicates 
existence of stable structures in the circumstellar disk, caused by presence of either 
low-mass secondary or a protoplanet [8,15].

The most uniform time-series of cTTS and HAeS were obtained in course of the ROTOR project
[16] in 1983-2004 at the Majdanak Observatory, Uzbekistan. The time series of 10-20 years
provide a rich database to study variability on time scales of 20 to 200 days, corresponding 
to Keplerian periods near the inner radius of the accretion disks. In this paper we report
on search for such periodicities in light variations of cTTS and HAeS.
\\
\\
\noindent {\bf 2. Observational data and methods of analysis}
\\
\\
\noindent 
The observational data were described by Grankin et al. [6] and are available in electronic form 
from CDS, Strasbourg (anonymous ftp to sdsarc.u-trasbg.fr). The observations of HAeS before
1999 were published in [17]; later observations are not published yet. All the data were reduced 
to the Johnson's UBVR system. For stars brighter than V=12 mag the standard deviation is 0.01 mag in
BVR and 0.05 mag in U. From the list of targets in the catalogue we selected those objects
which were observed during at least five seasons (maximum 23 seasons), with duration of
one season from 3 to 6 months. Usually, each target was observed once in every clear night
during a season. Typical light-curves of HAes and cTTS are shown in Fig.1.

%========================= Fig 1 ==========================
\begin{figure}
\begin{center}
\resizebox{12cm}{!}{\includegraphics{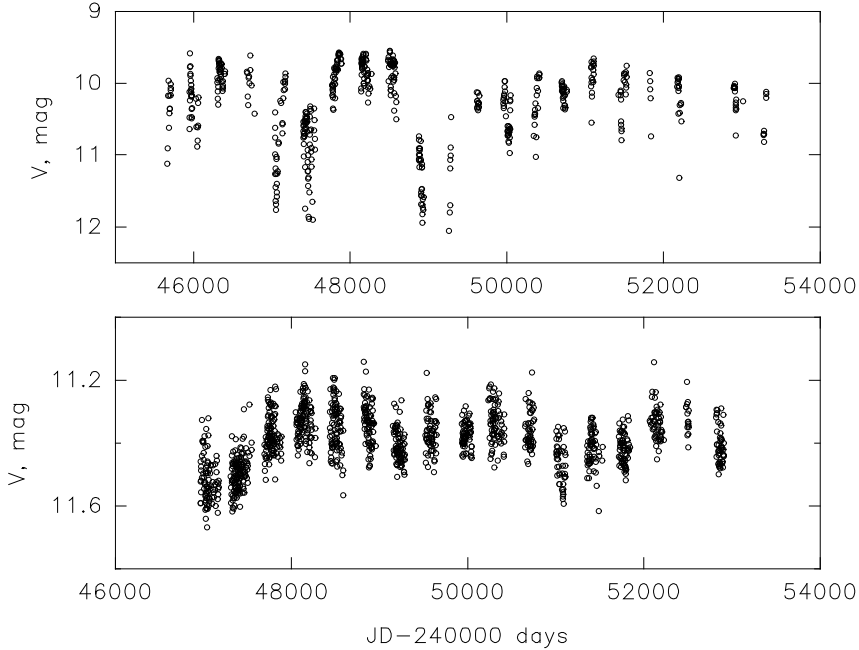}}
\caption{Typical light-curves of HAes and cTTS on time interval of about 20 years.
Upper panel: BF Ori, spectral type A2. Lower panel: DI Cep, spectral type G8.}
\label{fig:petrf1}
\end{center}
\end{figure}
%=========================================================

In this paper we analyse 11 HAeS showing effect of UX Ori, and 28 cTTS. All the 
light-curves of these stars show both the short term nigh-to-night variations and the long 
term variations from season to season. We are interested in probable periods on time scale
from 20 to 200 days, therefore variations on longer time scales are considered as a 
low-frequency trend. In order to remove the trend we subtracted the seasonal average from the 
data within each season. The light variations may be caused by both the periodical and 
stochastic processes, therefore we used two different methods: the digital spectral analysis
and the method of auto-correlation functions.
\\
\\
\noindent{\bf 3. Digital spectral analysis}
\\
\\
\noindent
In order to reveal the hidden periodic processes from the light curves we used different
methods of the power spectra estimations: CLEAN [18], Chi-square [19], SCARGLE [20] and the classical 
correlogram method CORRPSD [21]. The statistical characteristics of the power spectra
calculated by these methods appeared about the same, not counting small differences in the form,
therefore in the following we show and discuss only the results from the CORRPSD method.

From the point of view of the spectral analysis theory, our light curves represent discrete, 
unevenly sampled and rather short time sequences of data with significant seasonal gaps.
These properties result in badly predicted distortions of the corresponding estimates of the
power spectra and makes it impossible to reconstruct the true (consistent) power spectra.
In other words, the estimate of a power spectrum appeared to be unstable or inconsistent. 
For reconstruction of consistent estimate we used the method of smoothing the spectral estimate, 
as was first suggested in [22] for the case of correlogram method. Detailed discussion of the 
problems of reconstruction of consistent power spectrum in case of real astronomical time-series 
can be found in [23].

We used the smoothing algorithm described in [24]. The smoothing of power spectrum
was done by change in the correlation window width. Narrowing the correlation window leads 
to smaller variance of spectral estimate at the expense of spectral resolution.
The problem is to find a compromise between the spectral resolution and the consistency
of spectral estimate. We use the classical CORRPSD method and the Tukey window function.
In this method, in computation of a power spectrum one uses not the values of time series
but a finite sequence of auto-correlation function.

The Fig.2 shows typical power spectra of HAeS and TTS. Unlike HAeS, power spectra of many TTS
contain peaks corresponding to the periods of axial rotation ($<$10 days). The rotational modulation 
of light of cTTS is due to the cold and hot spots on their surface. We reserve the analysis of 
the rotational periods and physical characteristics of the spots for
another paper. Here we consider processes of lower frequencies, with characteristic time $>$10 days.
It is known that light variations of HAeS is usually slower than that of TTS [25]. 
This can be noticed from the shape of the power spectra in Fig.2: in HAeS the power is rising 
steeply towards the lower frequencies, while in TTS the spectrum is more flat. 
Let's express this quantitatively. By analogy with the energy distribution
in optical spectrum of a star, one can apply the method of colorimetry to the power spectrum.
The summed power within selected frequency bands can be measured, and the colour indexes can
be determined. We selected two bands: the low-frequency band from 0.005 to 0.0166 1/day
(periods from 60 to 200 days), and the mid-frequency band from 0.0166 to 0.01 (periods
from 60 to 200 days). The boundaries of these two bands are marked by dashed lines in Fig.2
Let's denote the summed power within the low-frequency band as LF, and  the summed power within 
the mid-frequency band as MF. Then, the ratio LF/MF is analog of the red colour index:
the larger LF/MF, the longer the characteristic time of variability. In case there is a periodical 
component in light variations, the LF/MF ratio depends on the value of the period.
\\
\\
%========================= Fig 2 ==========================
\begin{figure}
\begin{center}
\resizebox{12cm}{!}{\includegraphics{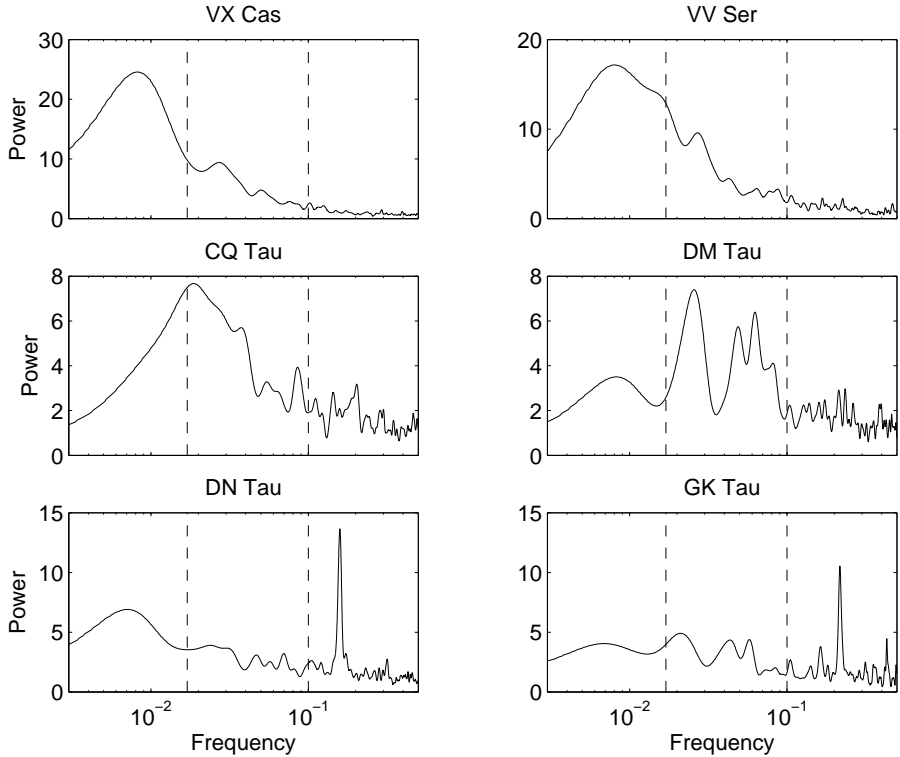}}
\caption{Examples of power spectra of HAeS (VX Cas and VV Ser) and cTTS (CQ Tau, DM Tau, DN Tau
and GK Tau). The dashed lines mark the frequency intervals used for determination of 
the colour index of the power spectra.}
\label{fig:petrf1}
\end{center}
\end{figure}
%=========================================================
\noindent{\bf 4. Auto-correlation functions} 
\\
\\
\noindent
Variability of a star can be characterised also by means of auto-correlation function
(ACF). When calculating the ACF, we remove the seasonal trend by the same way as 
in calculation of power spectrum. Besides of that, in those cases when a period of axial 
rotation is clearly present in the brightness variations, the periodical component
was also removed from the data set before calculating the ACF. We calculated the ACF from
all the light-curves in B, V and R bands. All the three ACFs are about the same, so in the
following we discuss only the ACF calculated for the V band.  As an example, ACFs for 
DR Tau (cTTS) and WW Vul (HAeS) at the time interval of 30 days are shown in Fig. 3.
Apparently, the widths of the central peak of these ACFs are different, which indicates
the different rates of brightness variations.

%========================= Fig 3 ==========================
\begin{figure}
\begin{center}
\resizebox{10cm}{!}{\includegraphics{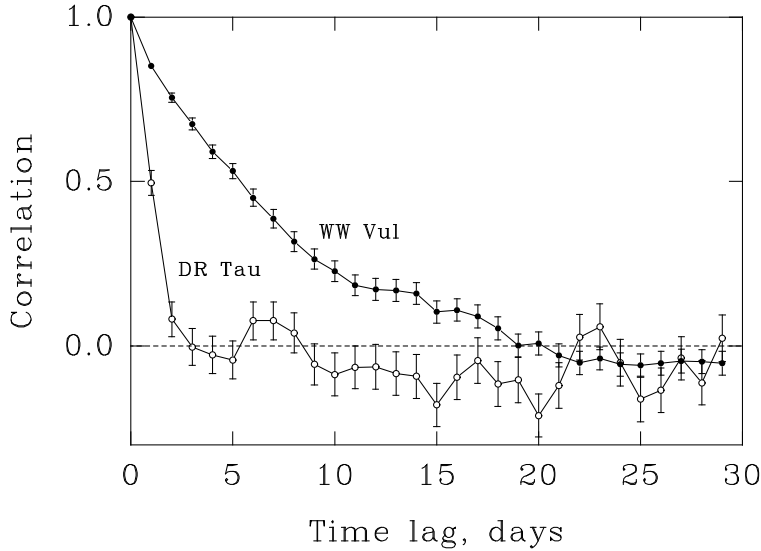}}
\caption{Auto-correlation functions of light curves of HAeS WW Vul and cTTS DG Tau}
\label{fig:petrf1}
\end{center}
\end{figure}
%=========================================================

Lets denote the width of ACF at zero level as W0. This parameter, sometimes called
''interval of correlation'', is the time interval during which stellar brightness changes 
from mean level to maximum or minimum. In case of sinusoidal variations with period P
this interval equals to 1/4 P. In case of non-periodical variations, W0 can be considered 
as a characteristic time of variability. In the following we will use the term 
''characteristic time'' in this meaning.
\\
\\
\noindent {\bf 5. Relationship between the characteristic time and luminosity}
\\
\\
\noindent
The measured values of LF/MF and W0 are given in Table 1, together with errors in W0. 
The errors in LF/MF is typically less than 1\%. Both parameters characterise
the same property: the mean rate of light variability, or the characteristic time of
light variability. The average value of W0 is 8.7 days for the group of cTTS, and 21.1 days
for the group of HAeS.

%========================= Fig 4 ==========================
\begin{figure}
\begin{center}
\resizebox{14cm}{!}{\includegraphics{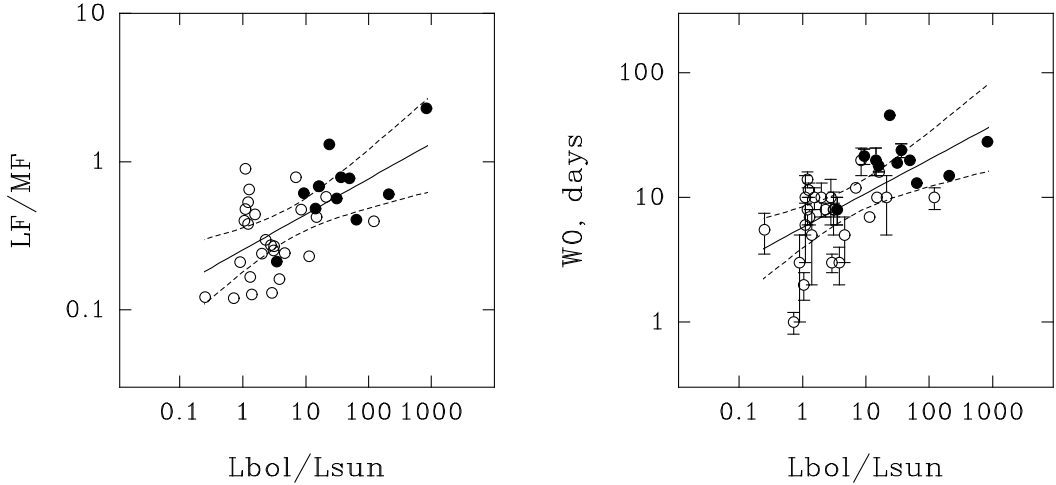}}
\caption{Correlation between the characteristic time of light variability and the total
bolometric luminosity of the 'star + disk' system. Open circles: HAeS; filled circles: cTTS}
\label{fig:petrf1}
\end{center}
\end{figure}
%=========================================================

When considering global stellar parameters, like mass and luminosity, we discovered 
dependence of both LF/MF and W0 from the {\it total bolometric luminosity} of the system 
''star + disk'' (Fig.4). This luminosity, denoted as L$_{tot}$, includes the bolometric luminosity 
of the star and the luminosity caused by accretion, from optical to IR down to 100 mkm.
the total bolometric luminosities and the corresponding references [26,27,29,35,36] are given in 
columns 2 and 3 in Table 1. The Fig.4 shows that both cTTS and HAeS groups comprise
a common dependence:  the larger the luminosity is, the slower the variability. The range of
2--3 dex in luminosity corresponds to the range of 1 dex in the characteristic time of variability. 
Probably, the characteristic time of variability is related to the disk size.

For the last decade, large progress was achieved in measurements of the inner radii 
of accretion disks by interferometric methods (see, e.g. review [4]). In the near-IR,
the spatial resolution of modern interferometers reaches 0.007 arcsec, which corresponds 
to 0.1 AU at the distance of 140 pc. The interferometric observations showed that the inner 
radius R$_{in}$ of accretion disk, where the dust sublimates, is $\sim$0.1 AU in cTTS and $\sim$0.5 AU in HAeS.
The values of R$_{in}$ for the stars under consideration and the references [28,30-34,37,38] are given in columns 
4 and 5 in Table 1.

%%%%%%%%%%%%%%%%%%%%%%%%%%%%%%%%%%%%%%%%%%%%%%%%%%%%%%%%%%%%%%%%%%%%%%%%%%%%%%%%%%%%%
\begin{table}
\begin{center}
\caption{Total bolometric luminosity, inner disk radius, and parameters of photometric variability}
\bigskip
\tabcolsep=2.5mm
\begin{tabular}{|r|r|r|r|r|r|r|}
\hline
    Object &       L$_{tot}$ (L$_\odot$) & Reference &  R$_{in}$ (AU) & Reference &  $W0$ &  $LF/MF$ \\
\hline
     1&  2& 3&  4&  5&  6& 7\\     
\hline
    AA Tau &       1.05 &    [26,27] &      0.084 &       [28] &        $2.0\pm0.5$ &      0.399\\
    AS 205 &       11.4 &       [29] &       0.140 &       [30] &       $7.0\pm0.2$ &      0.230 \\
    BP Tau &       1.25 &    [26,27] &      0.086 &       [31] &       $11.5\pm1.5$ &      0.649 \\
    BM And &          - &          - &      0.249 &       [28] &       $20.0\pm5.0$ &          - \\
    CI Tau &       1.55 &    [26,27] &      0.097 &       [28] &       $10.0\pm2.0$ &      0.440 \\
    DF Tau &       2.93 &  [26,27,29] &       0.090 &       [32] &       $3.0\pm0.5$ &      0.131 \\
    DG Tau &        6.90 & [26,27,29] &      0.142 &       [31] &      $12.0\pm0.3$ &      0.785 \\
    DI Cep &        8.50 &       [29] &      0.165 &       [28] &      $20.0\pm5.0$ &      0.475 \\
    DI Tau &        0.90 &    [26,27] &     (0.10) &          - &       $3.0\pm2.0$ &      0.210 \\
    DK Tau &        2.80 &    [26,27] &     (0.13) &          - &      $10.0\pm4.0$ &      0.273 \\
    DL Tau &        1.40 &    [26,27] &     (0.10) &          - &       $5.0\pm3.0$ &      0.127 \\
    DM Tau &       0.25 &       [26] &     (0.08) &          - &        $5.5\pm2.0$ &      0.122 \\
    DN Tau &        1.20 &    [26,27] &       0.070 &       [32] &      $8.0\pm2.0$ &      0.381 \\
    DR Tau &        3.80 & [26,27,29] &      0.103 &    [32,33] &       $3.0\pm1.0$ &      0.161 \\
    DS Tau &        1.10 &    [26,27] &     (0.10) &          - &       $6.0\pm3.0$ &      0.479 \\
    GG Tau &        2.30 &    [26,27] &     (0.12) &          - &       $8.0\pm1.0$ &      0.297 \\
    GI Tau &        1.20 &       [26] &     (0.10) &          - &      $14.0\pm2.0$ &      0.533 \\
    GK Tau &          2.00 &    [26,27] &     (0.12) &          - &    $10.0\pm3.0$ &      0.240 \\
    GM Aur &        1.10 &    [26,27] &      0.221 &       [31] &      $10.0\pm5.0$ &      0.893 \\
    GW Ori &        121 &       [29] &     (0.55) &          - &       $10.0\pm2.0$ &      0.397 \\
   LkCa 15 &       0.72 &    [26,29] &      0.099 &       [31] &        $1.0\pm0.2$ &      0.120 \\
  RW AurAB &       3.15 &    [26,29] &       0.130 &    [28,31] &       $8.0\pm1.0$ &      0.269 \\
    RY Tau &       16.3 & [26,27,29] &        0.300 &       [34] &     $16.0\pm1.0$ &      0.683 \\
    SU Aur &         15.0 &    [26,27] &       0.240 &    [32,33] &    $10.0\pm1.0$ &      0.422 \\
     T Tau &       21.1 & [26,27,29] &     (0.23) &          - &       $10.0\pm5.0$ &       0.580 \\
  UX TauAB &        1.30 &       [26] &     (0.10) &          - &       $7.0\pm1.0$ &      0.167 \\
    UY Aur &        3.10 &       [26] &       0.100 &       [32] &      $8.0\pm3.0$ &      0.251 \\
 V1121 Oph &        4.60 &       [29] &     (0.15) &          - &       $5.0\pm2.0$ &      0.242 \\
\hline
    BF Ori &        9.40 &       [35] &     (0.18) &          - &      $21.5\pm3.0$ &      0.612 \\
    BH Cep &       14.4 &       [36] &     (0.20) &          - &       $20.0\pm5.0$ &      0.484 \\
    CQ Tau &        3.50 &       [36] &       0.230 &       [37] &      $8.0\pm2.0$ &      0.212 \\
    KK Oph &         16.0 &    [29,35] &     (0.20) &          - &     $18.0\pm2.0$ &      0.683 \\
  LkHa 234 &        829 &    [29,35] &     (1.00) &          - &       $28.0\pm2.0$ &      2.298 \\
    RR Tau &         64.0 &       [36] &     (0.45) &          - &     $13.0\pm1.0$ &      0.406 \\
    SV Cep &         24.0 &       [36] &     (0.30) &          - &     $46.0\pm2.0$ &      1.113 \\
    UX Ori &         31.0 &       [36] &       1.090 &       [38] &    $19.0\pm1.0$ &      0.563 \\
    VV Ser &        208 & [29,35,36] &     (0.47) &       [37] &       $15.0\pm1.0$ &      0.603 \\
    VX Cas &         50.0 &       [36] &     (0.40) &          - &     $20.0\pm2.0$ &      0.773 \\
    WW Vul &       36.3 &       [36] &       0.990 &       [38] &      $24.0\pm3.0$ &      0.785 \\
\hline
\end{tabular}
\end{center} 
\end{table}
%%%%%%%%%%%%%%%%%%%%%%%%%%%%%%%%%%%%%%%%%%%%%%%%%%%%

In [31], a relationship between R$_{in}$ and the bolometric luminosity of the system ''star + disc''
was established. Using this relationship, we find dependence of the characteristic time of
variability W0 from R$_{in}$. Then, using stellar masses, we find a relationship between W0 
and the Keplerian period at the radius R$_{in}$. For some stars, where R$_{in}$ is not known, we 
estimate it from the known luminosity L$_{tot}$ and the relationship R$_{in}$ vs L$_{tot}$ [31]: these
estimated R$_{in}$ are given in brackets in Table 1.

The empirical relationships in Fig.4, although statistically significant, show rather 
large scatter of points. Therefore, in the following we analyse only mean values
(L$_{tot}$, R$_{in}$, M, P$_{kepl}$, W0) calculated for the two groups: cTTS and HAeS. These mean values and
the standard deviations of the mean are given in Table 2. It turns out, that the characteristic 
time of variability W0 is about 1/4 of the Keplerian period at the inner radius of accretion disk,
more precisely -- at the radius of dust sublimation (see the last two lines in Table 2.).
Apparently, the light variability is related to some processes at the inner radius of the 
dusty disk. 

This is not a trivial result, because the characteristic time of variability is measured from
the light curves without any model assumptions, while the Keplerian period is derived from
the spectral, photometrical and interferometrical observations and involves the evolutionary 
tracks and disk models.
\\

%%%%%%%%%%%%%%%%%%%%%%%%%%%%%%%%%%%%%%%%%%%%%%%%%%%%%%%%%%%%%%%%%%%%%%%%%%%%%%%%%%%%
\begin{table}
\begin{center}
\caption{Average parameters of cTTS (28 objects) and HAeS (11 objects)}
\bigskip
\tabcolsep=2.5mm
\begin{tabular}{|c|c|c|}
\hline
    Parameter         &    cTTS       &    HAeS \\
\hline
L$_{bol}$ (L$_\odot$) &  $8.8 ± 4.3$  & $116.9 ± 73.2$ \\
R$_{in}$(AU)        & $0.15 ± 0.02$ & $0.50 ± 0.11$ \\
M (M$_\odot$)         & $0.81 ± 0.14$ & $3.10 ± 0.4$ \\
\hline
P$_{kepl}$(d)         & $25.6 ±  2.9$ & $86.5 ± 27.3$ \\
$4*W0$ (d)            & $32.6 ±  3.4$ & $81.4 ± 10.2$ \\
\hline
\end{tabular}  
\end{center}
\end{table}

%%%%%%%%%%%%%%%%%%%%%%%%%%%%%%%%%%%%%%%%%%%%%%%%%%%%%%%%%%%%%%%%%%%%%%%%%%%%%%%%%%%%%%%

\noindent {\bf 6. Search for stable Keplerian periods}
\\
\\
\noindent
The obtained results give hopes that Keplerian periods can be detectable in the photometric
time series. Using the smoothed power spectra, described in Section 3, we composed a list
of all the periodic processes with a confidence level above 80\%. The periods caused by orbital
rotation of low-mass bodies must be stable in time, appearing in different seasons of observations. 
In order to select such stable periods we inspect phase diagrams for each season individually. 
In case the periods reveals itself in at least five seasons, we assumed the period as a stable one. 
As an example, the Fig.5 shows phase diagrams for the star AS 205 for period P1=24.8 days 
(the period of axial rotation of the star is 6.78 days). 
The period P1 can be noticed in the phase diagrams of 1987, 1988, 1989,
1992 and 1993 seasons. Note, that the time interval from 1987 to 1993 covers 88 periods.

%========================= Fig 5 ==========================
\begin{figure}
\begin{center}
\resizebox{16cm}{!}{\includegraphics{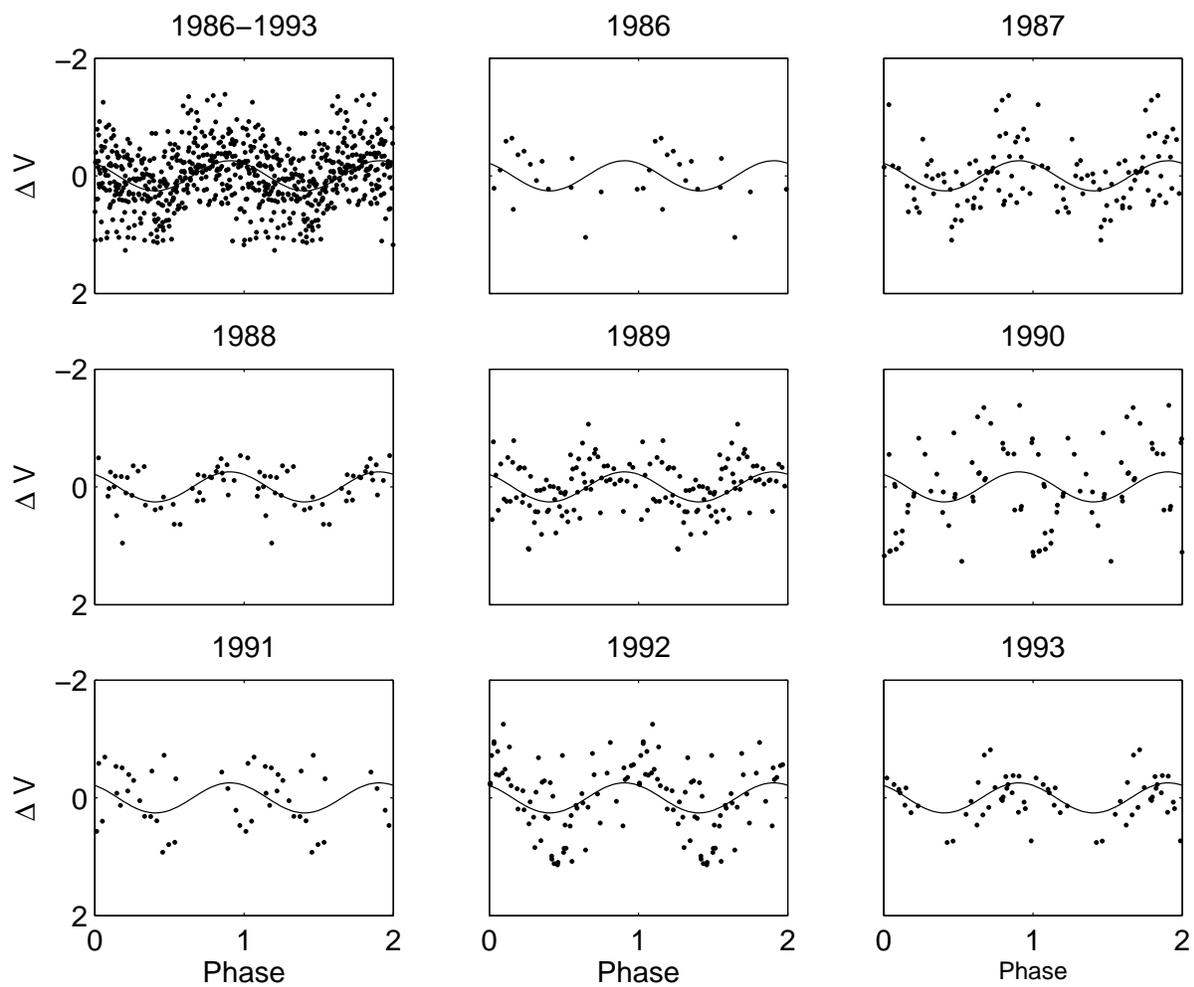}}
\caption{Seasonal light curves of AS205, convolved with period P=24.8 days. 
Initial epoch T0=2445000 is the same in each season. The sinusoid fit to the whole 
set of data is shown in each box}
\label{fig:petrf1}
\end{center}
\end{figure}
%=========================================================

Such relatively stable periods were found only in five cTTS: AS 205, CI Tau, DI Cep, GI Tau 
and GW Ori. Assuming that the periods are caused by orbital motion of bodies (protoplanets)
within the protoplanetary disks, and using the stellar masses, we estimated semi-major axes 
of the orbits. Table 3 lists the stars with relatively stable periods: there are star name,
spectral type, effective temperature, stellar luminosity, stellar mass, total number of 
observations (N1), number of seasons (N2), the period (P) and semi-major axis of the orbit 
corresponding to the period P.  

DI Cep was observed from 1986 to 2003; the period given in Table 3 was detected in data of 1986, 1989, 1991, 
1993, 1994 and 1999 seasons, which covers time interval of 40 cycles. GI Tau was observed from 1986 to 1992; 
the period was detected in 1987, 1988, 1989, 1990 and 1992, which covers 28 cycles. GW Ori was observed 
from 1987 to 1998 and the period was detected in 1988, 1989, 1992, 1993 and 1998, which covers 45 cycles. 
Note, that in this paper we do not consider the periods of axial rotation, which are detectable in light 
curves of many cTTS. The periods of axial rotation are shorter than Keplerian periods and are also quite 
stable.

Among the HAeS listed in Table 1, we did not find stable periodical variations of light. The long-term cyclic 
variations with a period of about 650 days are present in SV Cep; 
this was reported earlier in [12].
\\
\\
\noindent{\bf 7. Discussion and conclusions}
\\
\\
\noindent
In this work we analysed long-term UBVR photometry of cTTS and HAeS. We discovered relationship between the 
characteristic time of variability and the bolometric luminosity of the system "star + disk": the larger the 
luminosity is, the slower the variability.
The relationship is valid over a wide range of masses and luminosities from cTTS to HAeS.
Taking into account the known relationship between the inner radius of dusty disk R$_{in}$
and the bolometric luminosity, one may suggest that the characteristic time of 
photometric variability is related to some processes near the inner boundary of
the dusty disk. Using the R$_{in}$, the mass of the star and the 2nd Kepler's law, we estimated
mean Keplerian periods for 28 cTTS and 11 HAeS. It turns out that the observed characteristic time of 
variability is, on average, equal to 1/4 of the Keplerian period at the radius of 
dust sublimation.

By means of periodogram analysis of the light curves we found stable periods between 25 and 120 days in five 
cTTS. This implies existence of stable structures (protoplanets, dust clouds or density waves) within the 
circumstellar disks, rotating at Keplerian orbits at the distances of 0.1 to 0.5 AU from the star. This is in 
good agreement with the inner radii of the dusty disks of cTTS  known from IR interferometry.  
Two different mechanisms can be involved in interpretation of these results: 1) circumstellar extinction and 
2) accretion. In the first case, motion of a protoplanet causes density waves in 
the disk and in the disk wind. In the presence of dust in the disk it results in variability
of circumstellar extinction, provided a low inclination of light of sight to the disk plane.
This mechanism was considered by Tambovtseva and Grinin [43], who argued the ability of dust 
to survive in the disk winds of cTTS. In the second case, motion of a protoplanet in an eccentric orbit
can modulate the rate of accretion and, consequently, the stellar brightness. In this case the light 
variability can be observed at any inclination of line of sight to the disk plane.
The periodic modulation of accretion rate was observed in DQ Tau [44] and V4046 Sgr [45].
Similar scenario was also discussed in [17].
There is also a possibility of a ''hybrid'' scenario: increase of accretion rate is accompanied by
increase of mass-loss, which, in turn, leads to increase in circumstellar extinction. This 
mechanism was discussed in [46] with reference to V1184 Tau.

One of the most interesting case is the cTTS AS 205A (= V866 Sco). The star shows two distinct 
periods in light variations, 6.78 and 24.78 days, that persist over several years. The shortest 
period is due to axial rotation of the star. If the 24.78 days period is due to orbital motion of 
a gravitating object, there must be also variations in radial velocity of the star.
With the mass of the primary $\approx$ 1 M$_\odot$, its orbital velocity is expected to be above 5 km\,s$^{-1}$ 
in case the secondary is of stellar mass; from 1 to 5 km\,s$^{-1}$ in case of a brown dwarf, and below 
1 km\,s$^{-1}$ in case of a planet. The expected amplitude of radial velocity depends on inclination 
of the orbit to the line of sight, which can be estimated from the observed $v\,\sin i$, stellar radius
and the period of axial rotation (6.78 days). At present, only a few measurements of radial velocity
of AS 205 are available, allowing to suspect variability of radial velocity [47]. Note, that if there 
are spots on stellar surface, variations of radial velocity (within $v\,\sin i$) may appear also
due to distortions in photospheric line profiles [48].

Special spectral monitoring of the cTTS, listed in Table 3, is needed to clarify the nature 
of the periodic brightness variations. 
\\
\\
%%%%%%%%%%%%%%%%%%%%%%
{\bf References}
\\
\\
\noindent
1.  P. P. Petrov, Astrophysics 46, 506 (2003).

\noindent
2. L.~B.~F.~M.~Waters, C.~Waelkens, Ann. Rev. Astron. Astrophys. \textbf{36}, 233 (1998).

\noindent
3. C.~A.~Grady, {\it The Nature and Evolution of Disks Around Hot Stars}, Johnson City, Tennessee, 

USA (2004). Edited by R. Ignace and K.~G.~Gayley., ASP Conf. Series \textbf{337}, 155 (2005).

\noindent
4.~Millan-Gabet, F.~Malbet, R.~Akeson et al., Protostars \& Planets V, Big Island, Hawaii (2005), 

Edited by B.~Reipurth, D.~Jewitt, and K.~Keil (eds.), Univ. of Arizona Press, p.539 (2007).

\noindent
5.~P.~Grinin, N.~N.~Kiselev, G.~P.~Chernova et al., Astrophys. Space Sci. \textbf{186}, 283 (1991).

\noindent
6. K.~N.~Grankin, S.~Yu.~Melnikov, J.~Bouvier et al., Astron. Astrophys. \textbf{461}, 183 (2007).

\noindent
7. K.~N.~Grankin, J.~Bouvier, W.~Herbst, S.~Yu.~Melnikov, Astron. Astrophys. \textbf{479}, 827 (2008).

\noindent
8.  V. P. Grinin, L. V. Tambovtseva, N. Ya. Sotnikova, Astron. Lett. 30, 694 (2004).

\noindent
9. V.~P.~Tsessevich, V.~A.~Dragomiretskaya, {\it RW Aur type stars} (Kiev: Naukova Dumka, 1973).

\noindent
10. V.~S.~Shevchenko, K.~Grankin, M.~Ibragimov et al., Astrophys. Space Sci. \textbf{202}, 121 (1993).

\noindent
11. V. P. Grinin, A. N. Rostopchina, D. N. Shakhovskoi, Astron. Lett. 24, 802 (1998).

\noindent
12. A. N. Rostopchina, V. P. Grinin, D. N. Shakhovskoi, Astron. Lett. 25, 243 (1999).

\noindent
13. A. N. Rostopchina, V. P. Grinin, D. N. Shakhovskoi et al., Astron. Rep. 44, 365 (2000)

\noindent
14. C.~Bertout, Astron. Astrophys. \textbf{363}, 984 (2000).

\noindent
15. D. N. Shakhovskoi, V. P. Grinin, A. N. Rostopchina, Astrophysics 48, 135 (2005).

\noindent
16. V. S. Shevchenko, Herbig Ae/Be stars (Tashkent, Izdatel'stvo Fan, 1989, In Russian).

\noindent
17. W.~Herbst, V.~S.~Shevchenko, Astron.~J. \textbf{118}, 1043 (1999).

\noindent
18. D.~H.~Roberts, J.~Lehar, J.~W.~Dreher, Astron.~J. \textbf{93}, 968 (1987).

\noindent
19. K.~Horne, R.~A.~Wade, P.~Szkody, MNRAS \textbf{219}, 791 (1986).

\noindent
20. J.~H.~Horne, S.~L.~Baliunas, Astrophys.~J. \textbf{302}, 757 (1986).

\noindent
21. S. L. Marple, {\it Digital spectral analysis with applications} (Englewood Cliffs, NJ, 

Prentice-Hall, Inc., 1987).

\noindent
22. R.~B.~Blackman, J.~W.~Tukey, {\it The measurement of power spectra  from the} 

{\it point of view of communication engineering}, (New York: Dover Publications, Inc., 1958).

\noindent
23. V. Yu. Terebizh, {\it Introduction to Statistical Theory of Inverse Problems} 

(Moscow, FIZMATLIT, 2005).

\noindent
24. G. M. Jenkins, D. G. Watts, {\it Spectral analysis and its applications} 

(Holden-Day Series in Time Series Analysis, London: Holden-Day, 1969).

\noindent
25. W.~Herbst, D.~K.~Herbst, E.~J.~Grossman, Astron.~J. \textbf{108}, 1906 (1994).

\noindent
26. S.~J.~Kenyon, L.~Hartmann, Astrophys.~J. Suppl. Ser. \textbf{101}, 117 (1995).

\noindent
27. S.~Cabrit, S.~Edwards, S.~E.~Strom, K.~M.~Strom, Astrophys. J. \textbf{354}, 687 (1990).

\noindent
28. J.~A.~Eisner, L.~A.~Hillenbrand, R.~J.~White et al., Astrophys.~J. \textbf{669}, 1072 (2007).

\noindent
29. F.~Hamann, S.~E.~Persson, Astrophys.~J. \textbf{394}, 628 (1992).

\noindent
30. J.~A.~Eisner, L.~A.~Hillenbrand, R.~J.~White et al., Astrophys.~J. \textbf{623}, 952 (2005).

\noindent
31. R.~L.~Akeson, A.~F.~Boden, J.~D.~Monnier et al., Astrophys.~J. \textbf{635}, 1173 (2005).

\noindent
32. J.~Muzerolle, N.~Calvet, L.~Hartmann, P.~D'Alessio, Astrophys.~J. \textbf{597}, L149 (2003).

\noindent
33. R.~L.~Akeson, C.~H.~Walker, K.~Wood et al., Astrophys.~J. \textbf{622}, 440 (2005).

\noindent
34. A.~A.~Schegerer, S.~Wolf, Th.~Ratzka, Ch.~Leinert, Astron. Astrophys. \textbf{478}, 779 (2008).

\noindent
35. M.~Corcaron, T.~P.~Ray, Astron. Astrophys. \textbf{331}, 147 (1998).

\noindent
36. J.~Hernandez, N.~Calvet, C.~Briceno et al., Astron.~J. \textbf{127}, 1682   (2004).

\noindent
37. J.~A.~Eisner, B.~F.~Lane, L.~A.~Hillenbrand et al., Astrophys.~J. \textbf{613}, 1049 (2004).

\noindent
38. J.~D.~Monnier, R.~Millan-Gabet, R.~Billmeier et al., Astrophys.~J. \textbf{624}, 832 (2005).

\noindent
39. P.~Hartigan, S.~Edwards, L.~Ghandour, Astrophys.~J. \textbf{452}, 736 (1995).

\noindent
40. N.~Calvet, J.~Muzerolle, C.~Briceno et al., Astron.~J. \textbf{128}, 1294 (2004).

\noindent
41. P.~Manoj, H.~C.~Bhatt, G.~Maheswar, S.~Muneer, Astrophys.~J. \textbf{653}, 657 (2006).

\noindent
42. F.~D'Antona, I.~Mazzitelli, Astrophys. J. Suppl. Ser. \textbf{90}, 467 (1994).

\noindent
43. L.~V. Tambovtseva, V.~P.~Grinin, 2008, Astron. Lett. \textbf{34}, 231 (2008).

\noindent
44. G.~Basri, Ch.~M.~Johns-Krull, R.~D.~Mathieu, Astron.~J. \textbf{114}, 781 (1997).

\noindent
45. H.~C.~Stempels, C.~F.~Gahm, Astron. Astrophys. \textbf{421}, 1159 (2004).

\noindent
46. V.~P.~Grinin, A.~A.~Arkharov, O.~Yu.~Barsunova et al., Astron. Lett. \textbf{35}, 114 (2009).

\noindent
47. C.~H.~F.~Melo, Astron. Astrophys. \textbf{410}, 269 (2003).

\noindent
48. H.~C.~Stempels, G.~F.~Gahm, P.~P.~Petrov, Astron. Astrophys. \textbf{461}, 253 (2007).
\\
\\
{\it Published in Astronomy Reports, 2010, V.54, P.163}

\end{document}